\documentclass[twocolumn,showpacs,preprintnumbers,amsmath,amssymb]{revtex4}

\usepackage{graphicx}
\usepackage{dcolumn}
\usepackage{bm}
\usepackage{amsthm}
\usepackage{graphicx}
\usepackage{enumerate}

\begin{document}


\title{Semi-loss-tolerant strong quantum coin-flipping protocol \\using quantum non-demolition measurement}

\author{Qian Yang$^{1,2,3}$, Jia-Jun Ma$^{1,3}$, Fen-Zhuo Guo$^{2}$\footnote{Email: gfenzhuo@bupt.edu.cn} and Qiao-Yan Wen$^{1}$}
\affiliation{ $^{1}$ State Key Laboratory of Networking and Switching Technology, Beijing University of Posts and Telecommunications, Beijing, 100876, China\\
        $^{2}$ School of Science, Beijing University of Posts and Telecommunications, Beijing, 100876, China\\
        $^{3}$ Center for Quantum Information, IIIS, Tsinghua University, Beijing, 100084, China}

\date{\today}

\begin{abstract}
In this paper, we present a semi-loss-tolerant strong quantum coin-flipping (QCF) protocol with the best bias of 0.3536. Our manuscript applies Quantum non-demolition (QND) measurement to quantum coin-flipping protocol. Furthermore, a single photon as a single qubit is used to avoid the difficult implementation of EPR resources. We also analyze the security of our protocol obtaining the best result among all coin-flipping protocols considering loss. A semi-loss-tolerant Quantum Dice Rolling (QDR) protocol is first proposed, and the security of corresponding three-party QDR is analyzed to better demonstrate the security of our QCF.

\end{abstract}

\pacs{03.67.Dd, 03.67.Hk, 89.70.-a}
\maketitle

\section{\label{sec:level1}INTRODUCTION}

{

Quantum Coin Flipping (QCF) is a cryptographic task first introduced by Blum in 1981\cite{M. Blum}. The goal of QCF is to allow two parties (often referred as Alice and Bob) who are distrustful and spatially separated to generate a random bit. While the value of this random bit cannot be controlled by anyone of them. The generalization of coin flipping is dice rolling (DR), which was extensively introduced in 1999 by Feige U\cite{Feige U} in classical settings. It is a cryptographic problem originally proposed by N Aharon and J Silman\cite{N Aharon and J Silman}, describing N remote distrustful parties must decide on a random string between 0 and $N-1$.

There are two variants of QCF: ``strong'' CF (SCF)\cite{Aharonov,Ambainis,Spekkens1,Colbeck,Chailloux and Kerenidis} and ``weak'' CF (WCF)\cite{Mochon1,Spekkens2,Mochon2}. In SCF neither party is aware of the other's preference for the coin's outcome, while in WCF the parties have opposite and known preferences. Obviously, every strong CF protocol can also be used to implement a weak CF protocol, but the converse statement is generally not true. The security of a CF protocol is quantified by the biases $\epsilon_A^{(i)}$and $\epsilon_B^{(i)} (i\in\{0,1\})$; if $P_A^{(i)^* }$ and $P_B^{(i)^* }$ are the maximal probabilities that a dishonest Alice or Bob can force the outcome to i, then
\begin{equation}
\epsilon_j^{(i)}=P_j^{(i)^* }-1/2, i \in 0,1,j=A,B.
\end{equation}

In classical settings, a dishonest party who is given unlimited computational power, can always fully bias the outcome as he or she wants, i.e., $\epsilon=1/2$\cite{J. Kilian}. In contrast, this is not the case in the quantum world: Unconditional secure coin flipping is possible. Although Mayers\cite{Mayers} and Lo Chau\cite{Lo and Chau}'s results implied perfect quantum coin flipping (the possibilities of both 0 and 1 are $1/2$ no matter what strategies a cheater uses) is impossible, it can be guaranteed that neither of the two parties can totally control the outcome (which is impossible by classical means). The first strong coin flipping protocol was proposed by Aharanov et al.\cite{Aharonov} with a bias of 0.414. Subsequently Ambainis\cite{Ambainis}, as well as Spekkens and Rudolph\cite{Spekkens1} independently improved this bound to 0.25. Unfortunately it was proven by Kitaev\cite{Kitaev} that no quantum strong coin flipping protocols can enjoy a bias less than 0.207 and this bound has been saturated by Chailloux and Kerenidis's protocol\cite{Chailloux and Kerenidis}. Compared with quantum SCF, quantum WCF is less well studied, Spekkens and Rudolph\cite{Spekkens2} first introduced a family of protocols with a bias of 0.207 and Mochon then improved it to 0.192 and finally to any $\epsilon\geq0$\cite{Mochon2}.

Although a lot of progress has been made along the way of exploring the least bias protocols, there is a common limit of previous results: practical issues were not taken into consideration. On imperfect practical conditions -such as losses and noise in the quantum channel as well as in the quantum memory storage- many protocols will be totally failed. Losses were first analyzed during the most common practical imperfection in the long distance communication. In 2008, Berl\'in \emph{et al}\cite{Berlin} (see also Ref.\cite{Nguyen}) implied a loss-tolerant SCF protocol with a bias of 0.4. The meaning of ``loss-tolerant'' here is defined by \cite{Berlin} when the protocol is impervious to any type of losses. After that Aharon et al.\cite{Aharon} announced a family of loss-tolerant quantum coin flipping protocols achieving a smaller bias than Berl\'in \emph{et al}. Very recently, Andre Chailloux\cite{Chailloux} added an encryption step to Berl\'in \emph{et al}'s protocol and introduced an improved loss-tolerant quantum coin flipping with bias 0.359. This result was slightly improved by Ma \emph{et al}\cite{Jia-Jun Ma} to 0.3536 with a semi-loss-tolerant strong coin-flipping protocol using EPR pairs. The meaning of ``semi-loss-tolerant'' here is explained by \cite{Jia-Jun Ma} when the protocol isn't impervious to certain types of losses and its security varies with the degree of those types of losses.

Compared with all the pratical SCF considering loss mentioned above, our protocol presented here obtains the best bias of 0.3536. Our manuscript applies Quantum non-demolition (QND) measurement to quantum coin-flipping protocol. Furthermore, we make use of the single photon as a single qubit avoiding the difficult implementation of EPR resources. After detailed security analysis, we find the security of both our protocol and three-party dice rolling protocol constructed by our protocol achieve the best result among all coin-flipping protocols considering loss.

After this introduction, the structure of the paper is organized as follows. We begin our protocol in Sec.~2 with a contrast to the protocol in \cite{Jia-Jun Ma}. In Sec.~3, We analyze the security of our protocol and better demonstrate it using an impressive three-party DR protocol, from which N-party DR protocol will also be of best bias. Conclusions and open problems are presented in Sec.~4.

\section{QND-BASED SEMI-LOSS-TOLERANT COIN-FLIPPING PROTOCOL}
{Recently the least bias among all SCF protocols considering loss is 0.3536 in \cite{Jia-Jun Ma}. And this protocol utilizes EPR pairs to make it come true. Whereas, quantum entanglement, as a physical resource, is of great difficulty to prepare in practice. Considering this, we try to give a more practical way to achieve the best result by utilizing single photon instead of EPR pairs. Correspondingly we change the protocol steps using Quantum nondemolition (QND) measurements\cite{Braginsky}. Here is our protocol.

\begin{enumerate}[1.]
\item  We say of $|\varphi(a,r_A)\rangle$ that $a$ is the basis and $r_A$ is the bit which could be showed as follows.
\begin{displaymath}
a=0\left\{
   \begin{aligned}
   |\varphi_{(0,0)}\rangle =|0\rangle\\
   |\varphi_{(0,1)}\rangle=|1\rangle \\
   \end{aligned}
   ,
   \right.
a=1\left\{
   \begin{aligned}
   |\varphi_{(1,0)}\rangle=\cos\alpha|0\rangle+\sin\alpha|1\rangle\\
   |\varphi_{(1,1)}\rangle=\sin\alpha|0\rangle-\cos\alpha|1\rangle \\
   \end{aligned}
   .
   \right.
\end{displaymath}
Alice prepares one state $|\varphi_{(a,r_A )}\rangle$ from {$|\varphi_{(0,0)}\rangle =|0\rangle,|\varphi_{(0,1)}\rangle=|1\rangle,|\varphi_{(1,0)}\rangle=\cos\alpha|0\rangle +\sin\alpha|1\rangle,|\varphi_{(1,1)}\rangle=\sin\alpha|0\rangle -\cos\alpha|1\rangle $}($0\leq\alpha\leq\pi/2$) with basis $a (0,1)$ and bit $r_A (0,1)$ chosen independently at random, then she sends the single photon to Bob.
\item Bob makes sure that he received this photon using QND measurements, keeps the received qubit in his quantum memory storage, and notices Alice about it. Otherwise, he will restart the protocol.
\item Bob sends Alice a randomly selected classical bit $b$.
\item Alice informs Bob of her selected single photon $|\varphi_{(a,r_A )}\rangle$.
\item Bob measures the qubit in the quantum memory according to Alice's announcing $a$. If he detects it, whose outcome is denoted as $r_B$, and finds that $r_A \neq r_B$, he aborts the protocol, calling Alice a cheater. If $r_A = r_B$ or even he doesn't detect the qubit due to the probability $p (0\leq p\leq1)$ that the qubit in Bob's quantum memory storage is lost, the outcome of the coin flipping is $b\oplus r_A$.
\end{enumerate}

In step 1, we just make use of a single photon instead of EPR pairs. Because of the QND measurement in step 2, Bob can justify whether the single photon arrives or not. Combining those two implements, we make it more feasible to realize the protocol with current technology.

The key difference between the protocol in \cite{Jia-Jun Ma} and ours is how to choose a method to better solve the problem that the qubit-receiver (Bob in our protocol) may receive no qubit so that the qubit-sender (Alice in our protocol) can announce any result to get the result she wants. The protocol in \cite{Jia-Jun Ma} chooses to utilize EPR pairs and our intention is to make it more practical by using one single photon with QND measurement. The remaining steps of the two protocols are equivalent.
}

\section{SECURITY ANALYSIS}
{
Let's begin with a comparison of security between three protocols: Ma \emph{et al}'s protocol\cite{Jia-Jun Ma}, Berl\'in \emph{et al}'s protocol\cite{Berlin} and ours. On one hand, we analyze the differences between the maximum bias of Alice, $\epsilon_{A}$. In Ma \emph{et al}'s protocol, $\epsilon_{A}^{(M)}=(1-p)\cdot\frac{\sin\alpha}{2}+p\cdot1/2$, $\frac{\sin\alpha}{2}$ is the probability that Alice's maximum bias when Bob successfully detect his particle with $1-p$ and this probability becomes $1/2$ (That is to say, Alice can always bias the result to what she wants with probability $1/2+1/2=1$)when Bob doesn't detect his particle in his quantum memory storage with $p$. Similarly, in Berl\'in \emph{et al}'s protocol, $\epsilon_{A}^{(B)}=1/2\cdot\frac{\sin\alpha}{2}+1/2\cdot1/2$ (The original text utilizes $\frac{1+2\alpha'\beta'}{4}$ where $\alpha'$ in the original text is $\cos\frac{\alpha}{2}$ here, and $\beta'$ is $\sin\frac{\alpha}{2}$ here, that is to say, $\frac{1+2\alpha'\beta'}{4}$ = $\frac{1+2\cos\frac{\alpha}{2}\sin\frac{\alpha}{2}}{4}$ = $1/2\cdot\frac{\sin\alpha}{2}+1/2\cdot1/2$), in which $\frac{\sin\alpha}{2}$ is the probability that Bob's measurement bases are the same as what Alice reveals and this probability turns to be $1/2$ when their measurement bases are not the same. The difference between $\epsilon_{A}^{(M)}$ and $\epsilon_{A}^{(B)}$ is because Ma \emph{et al}'s protocol manages to keep the two parties' bases consistent to decrease the bias of Alice.

When coming to $\epsilon_{A}^{(Y)}$ of our protocol, we always assume Bob is honest but Alice isn't. The only difference between Berl\'in \emph{et al}'s protocol and ours is that, we delay the second step in Berl\'in \emph{et al}'s protocol and let Bob measure his particle after Alice's announcement of measurement bases to keep the two parties' bases consistent just as what Ma \emph{et al} have done in their protocol. Then how could we guarantee the particle's arrival in the second step of Berl\'in \emph{et al}'s protocol, you may wonder? We utilize the QND measurement to make it come true. This is only related to Bob's action, so $\epsilon_{A}^{(Y)}$ isn't affected, that is, our result is the same as Ma \emph{et al}'s protocol: the maximum bias is $\frac{\sin\alpha}{2}$ if Bob successfully detect the single photon; it becomes $1/2$ if Bob doesn't. As a result,
\begin{equation}
\epsilon_{A}^{(Y)}=\epsilon_{A}^{(M)}=(1-p)\cdot\frac{\sin\alpha}{2}+p\cdot1/2.
\end{equation}

On the other hand, let's think about the difference between the maximum bias of Bob $\epsilon_{B}$. $\epsilon_{B}^{(M)}=\epsilon_{B}^{(B)}=\frac{\cos\alpha}{2}$, which demonstrates that Ma \emph{et al}'s protocol doesn't decrease the maximum bias of Bob. Similarly, the only cheating strategy of Bob in our protocol is the same as that of Bob in Berl\'in \emph{et al}'s protocol, that is, Bob uses an optimal measurement bases to measure and guess the single photon from Alice, then announcing a proper $b$ to get $b\oplus r_A$ which he wants. Thus,
\begin{equation}
\epsilon_{B}^{(Y)}=\epsilon_{B}^{(M)}=\epsilon_{B}^{(B)}=\frac{\cos\alpha}{2}.
\end{equation}

Consequently we get our $\epsilon_{A}^{(Y)}=(1-p)\cdot\frac{\sin\alpha}{2}+p\cdot1/2$ and $\epsilon_{B}^{(Y)}=\frac{\cos\alpha}{2}$ which is the same as those in Ma \emph{et al}'s protocol. The result is, both Alice and Bob, have an optimal cheating strategy capable of producing their desired outcome with 0.8536 probability of success (assuming the other player is honest).

Note that quantum coin-flipping protocol can be used to construct quantum dice rolling protocol. To better show the security of our protocol, we will analyze the security of three-party loss-tolerant dice rolling protocol constructed by our coin flipping protocol. Three-party semi-loss-tolerant dice rolling protocol is given as follows:
	
\emph{The first round} Alice and Bob roll the dice according to QND-BASED SEMI-LOSS-TOLERANT COIN-FLIPPING PROTOCOL described above. In the final step, if Bob detects the qubit, whose outcome is denoted as $r_B$, and finds that $r_A \neq r_B$, he aborts the protocol, calling Alice a cheater. If $r_A = r_B$ or even he doesn't detect the qubit, the outcome of the coin flipping is $b\oplus r_A$. Here we can suppose that Alice will win the first round if $b\oplus r_A$ is 0, and Bob will win the first round if $b\oplus r_A$ is 1. The winner is supposed to be Alice and without losing the generality will join the next competition.

\emph{The second round} Alice and Charlie roll the dice based on QND-BASED SEMI-LOSS-TOLERANT COIN-FLIPPING PROTOCOL described above. If Charlie detects it, whose outcome is denoted as $r_C$, and finds that $r_A \neq r_C$, he aborts the protocol, calling Alice a cheater. If $r_A = r_C$ or even he doesn't detect the qubit, the outcome of the coin flipping is $c\oplus r_A$. Here we can suppose that Alice will win the second round if $c\oplus r_A$ is 0, and Charlie will win the second round if $c\oplus r_A$ is 1. The winner is the final winner of the three parties.

According to the definition in \cite{N Aharon and J Silman}, DR protocol is fair if and only if
\begin{equation}
\overline{P_A^*}=\overline{P_B^*}=\overline{P_C^*},
\end{equation}
with $\overline{P_A^*}(\overline{P_B^*},\overline{P_C^*})$is the maximum probability that party A (B, C) loses. And we analyze the
following context based on the maximum probability of loss.

What's more, we will be interested in the N ``worst case'' scenarios to maximize the bias, where all but
one of the parties are dishonest and moreover, are cooperating with one another, using the classical and
quantum communication channels.

First, let's consider the maximum probability that party Alice loses, $\overline{P_A^*}$. According to the idea of N ``worst case'', we assume Bob and Charlie are dishonest and cooperating with each other while Alice is honest. So Alice will lose the first round with the probability of $\frac{1+\cos\alpha}{2}$ . Otherwise, if she wins the first round, she will lose the second found with the probability of $(1-\frac{1+\cos\alpha}{2})\cdot\frac{1+\cos\beta}{2}$ due to Charlie's maximum bias $\frac{\cos\beta}{2}$ in this round. As a result, the maximum probability that Alice loses is
\begin{equation}
\overline{P_A^*}=\frac{1+\cos\alpha}{2}+(1-\frac{1+\cos\alpha}{2})\cdot\frac{1+\cos\beta}{2}
\end{equation}
where $0\leq\alpha, \beta\leq\pi/2$ exist in the basis of each stage.

In the same way, we get the maximum probability that Bob loses
\begin{equation}
\begin{aligned}
\overline{P_B^*}&&=&\frac{1+p+(1-p)\sin\alpha}{2}\\
&&+&[1-\frac{1+p+(1-p)\sin\alpha}{2}]\cdot\frac{1+\cos\beta}{2},
\end{aligned}
\end{equation}
and the maximum probability that Charlie loses is
\begin{equation}
\overline{P_C^*}=\frac{1+p+(1-p)\sin\beta}{2}.
\end{equation}

On the whole, this protocol is fair iff (4). Solve those equations (4)(5)(6)(7) we get
\begin{equation}
\alpha=\arcsin\frac{-p(1-p)+ \sqrt{2-2p}}{2-2p+p^2}.
\end{equation}
\begin{equation}
\begin{aligned}
& & & [\frac{16}{(1-cos\alpha)^2}+\frac{4}{(1-p)^2}]\overline{P_C^*}^2\\
&- & & [\frac{4(1+p)}{(1-p)^2}+\frac{16}{1-cos\alpha}(\frac{1+cos\alpha}{1-cos\alpha}+\frac12)]\overline{P_C^*}
\\
& +& & 4(\frac{1+cos\alpha}{1-cos\alpha}+\frac12)^2+(\frac{1+p}{1-p})^2-1=0
\end{aligned}
\end{equation}
\begin{equation}
\beta=\arcsin\frac{2\overline{P_C^*}-1-p}{1-p}.
\end{equation}

As a result, we can obtain the maximal fair bias of our protocol  $\epsilon(p)=\overline{P_A^*}-\frac{2}{3}=\overline{P_B^*}-\frac{2}{3}=\overline{P_C^*}-\frac{2}{3}$ (a function of $p(0\leq p\leq1)$) which is showed in FIG.1. We can see it clearly in Curve Simulation(FIG. 1)$, \epsilon(p)$ monotonously decreases as p decreases. When $p$ approaches 0, the maximal fair bias of our protocol $\epsilon(p)$ is 0.2899 and the maximum probability that party Alice (Bob or Charlie) loses becomes 0.2899+0.6667=0.9566 which has also been the best result up to now in terms of QDR considering loss. Obviously, we can find that N-party QDR constructed by our protocol will be also of best bias. A compact six-round weak three-sided DR protocol is constructed in \cite{N Aharon and J Silman} using three-round weak imbalanced CF protocol in each two stages. However, it cannot be loss-tolerant. Our protocol could be more practical and at the same
time, more secure with a lower bias.

\begin{figure}
  \centering
  \includegraphics[width=8cm]{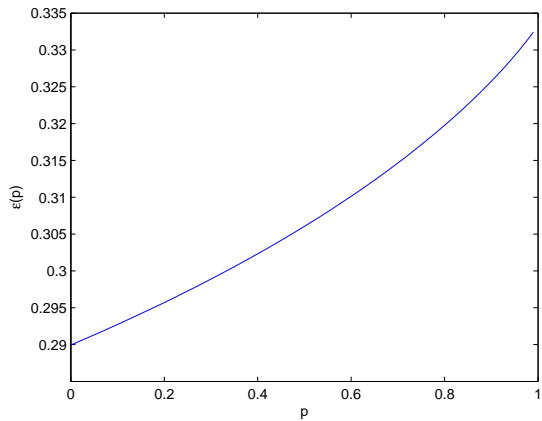}\\
  \caption{Maximal fair bias $\epsilon(p)$ is a function of $p$, it decreases with decreasing $p$, and our optimal bias achieves the best result.}
  \label{fig:example}
\end{figure}

}

\section{Conclusion}
{
To sum up, we get a semi-loss-tolerant strong quantum coin-flipping protocol using quantum non-demolition (QND) measurement, and the innovation points in our paper can be summarized as follows:

\begin{enumerate}[1.]

\item We utilize single photon to avoid the difficult implement of EPR resources, making our protocol more feasible in practice.

\item  We offer a new method to solve the problem that the qubit-receiver (Bob in our protocol) may receive no qubit so that the qubit-sender (Alice in our protocol) could announce any result to get the result she wants by combining QND measurement and the usage of single photon.

\item In terms of bias, the most important indicator of QCF, we obtain the best one of 0.3536 over all the QCF considering loss.

\item Quantum coin-flipping protocol can be used to construct quantum dice rolling protocol. To better analyze the security of our QCF, our manuscript first propose a semi-loss-tolerant QDR, and analyze the security of corresponding three-party QDR. When $p$ approaches 0, the maximal fair bias of our protocol $\epsilon(p)$ is 0.2899 and the maximum probability that party Alice (Bob and Charlie) loses becomes 0.2899+0.6667=0.9566 which has also been the best result up to now in terms of QDR considering loss.
\end{enumerate}

At the same time, a problem may emerge in the practical implementation because our bias is a function of parameter $p$ which is claimed by Bob. This is also demonstrated in\cite{Jia-Jun Ma}. At the same time, it is necessarily important that we continue to find a safer loss-tolerant quantum coin-flipping protocol with a smaller bias.
\begin{acknowledgments}
This work is supported by NSFC (Grant Nos. 61300181, 61272057, 61202434, 61170270, 61100203, 61121061), Beijing Natural Science Foundation (Grant No. 4122054), Beijing Higher Education Young Elite Teacher Project (Grant Nos. YETP0475, YETP0477). It is also supported by the National Basic Research Program of China Grant 2011CBA00300, 2011CBA00302, the National Natural Science Foundation of China Grant 61073174, 61033001, 61361136003.
\end{acknowledgments}}

\end{document}